\documentclass[aps,prl,twocolumn,superscriptaddress,longbibliography]{revtex4-1}
\usepackage{amsmath}
\usepackage{hyperref}
\hypersetup{
	colorlinks=true,
	linkcolor=blue,
	filecolor=blue,
	citecolor = blue,      
	urlcolor=cyan,
}

\usepackage{graphicx}
\usepackage{dcolumn}
\usepackage{bm}
\usepackage{natbib}
\usepackage[utf8]{inputenc}
\usepackage[T1]{fontenc}
\usepackage{mathptmx}
\usepackage{amsfonts}
\usepackage{amsmath}

\begin{document}

\title{Transforming space with non-Hermitian dielectrics}
\author{Ivor Kre\v{s}i\'{c}} \email{ivor.kresic@tuwien.ac.at} \email{ikresic@ifs.hr}
\affiliation{Institute for Theoretical Physics, Vienna University of Technology (TU Wien), Vienna, A–1040, Austria}
\affiliation{Institute of Physics, Bijeni\v cka cesta 46, 10 000 Zagreb, Croatia}

\author{Konstantinos G. Makris}
\affiliation{ITCP-Physics Department, University of Crete, Heraklion, 71003, Greece}
\affiliation{Institute of Electronic Structure and Lasers (IESL), Foundation for Research and Technology – Hellas, Heraklion, 71110, Greece}
\author{Ulf Leonhardt}
\affiliation{Department of Physics of Complex Systems, Weizmann Institute of Science, Rehovot 761001, Israel}
\author{Stefan Rotter}
\affiliation{Institute for Theoretical Physics, Vienna University of Technology (TU Wien), Vienna, A–1040, Austria}

\date{\today}

\begin{abstract}
Coordinate transformations are a versatile tool to mould the flow of light, enabling a host of astonishing phenomena such as optical cloaking with metamaterials. Moving away from the usual restriction that links isotropic materials with conformal transformations, we show how non-conformal distortions of optical space are intimately connected to the complex refractive index distribution of an isotropic non-Hermitian medium. Remarkably, this insight can be used to circumvent the material requirement of working with refractive indices below unity, which limits the applications of transformation optics. We apply our approach to design a broadband unidirectional dielectric cloak, which relies on non-conformal coordinate transformations to tailor the non-Hermitian refractive index profile around a cloaked object. Our insights bridge the fields of two-dimensional transformation optics and non-Hermitian photonics.
\end{abstract}

\maketitle

The introduction of coordinate transformations into optics has enabled a host of interesting applications such as optical cloaking—the ability of a device to conceal an object by shielding it from interacting with the incoming light \cite{leonhardt_optical_2006,pendry_controlling_2006}. However, the early designs of optical cloaks were hindered by demanding requirements, which prompted the search for materials with unconventional properties \cite{chen_transformation_2010}. Although this search continues today, some of these requirements, such as a vanishing refractive index, anisotropy or a non-vanishing magnetic susceptibility, have been found to come with severe restrictions such as a spectrally narrow optical response window. In spite of great efforts to circumvent these obstacles \cite{li_hiding_2008, valentine_optical_2009, gabrielli_silicon_2009,ergin_three-dimensional_2010,ni2015ultrathin,hsu2015extremely,mccall2018roadmap}, some applications like stand-alone cloaks \cite{leonhardt_optical_2006,pendry_controlling_2006, schurig_metamaterial_2006}, still rely on such properties. 

On another seemingly unrelated front, that of non-Hermitian photonics \cite{makris_beam_2008,lin_unidirectional_2011,chong_coherent_2010,longhi_parity-time_2017,el-ganainy_non-hermitian_2018,ozdemir_paritytime_2019,miri_exceptional_2019,shunyu2018}, engineering the imaginary part of the index of refraction has led to a plethora of experimental demonstrations \cite{ruter_observation_2010,regensburger_paritytime_2012,feng_experimental_2013,peng_paritytime-symmetric_2014,peng2014_2,feng_single-mode_2014,hodaei_parity-time-symmetric_2014,doppler_dynamically_2016,weidemann2020} with various novel applications \cite{assawaworrarit_robust_2017,pichler_random_2019,zhang2020,xia_nonlinear_2021}. Gain and loss provides an extra degree of freedom (different from negative or anisotropic electromagnetic responses) which offers an alternative route for controlling the flow  of light with a non-Hermitian medium.  

In this Letter we demonstrate how transformation optics and the engineering of non-Hermitian media can be directly linked. 
Specifically, in isotropic materials the non-conformal transformations naturally lead to spatially modulated gain and loss. When using them for constructing an invisibility cloak, such materials not only bend the phase fronts of light around the cloaked object, but the involved gain and loss distribution also provides sources and sinks for the probing light field. This can be used for 
designing a unidirectional broadband non-Hermitian version of the Zhukovsky cloak \cite{leonhardt_optical_2006,leonhardt_geometry_2010,xu_conformal_2015}. Rather than featuring anisotropic \cite{pendry_controlling_2006}, epsilon near zero (ENZ) \cite{alu2007epsilon,moitra2013realization} or negative index \cite{pendry_negative_2000,valentine2008three} materials, this non-Hermitian cloak just consists of a dielectric isotropic medium with spatially modulated gain and loss. With the underlying strategy to use non-conformal maps to design non-Hermitian index landscapes, we also open up new directions for the application of transformation optics in general. We envision, for example, that based on our results many of the existing conformal mapping setups in two-dimensional (2D) media~\cite{xu_conformal_2015} could also be considered for potential non-conformal extensions. The resulting non-Hermitian distributions can then be experimentally implemented with a spatially modulated pump beam~\cite{bachelard2012,hisch2013,bachelard2014}.

Our starting point is the Helmholtz equation that describes the scattering of a linearly polarized electric field of a given wavelength $\lambda_0$ at a 2D isotropic material landscape.  Following the strategy of transformation optics, we first consider a ``virtual space'' with coordinates ($x'$, $y'$), in which the incoming light sees a homogeneous medium with a constant and real refractive index $n_0$. The corresponding Helmholtz equation is given as follows: \begin{align}\label{eq:solvirt}
\Delta' E(x',y')+n_0^2\,k_0^2\,E(x',y')=0,
\end{align} 
where $\Delta'=\nabla'^2=\partial^2/\partial x'^2+\partial^2/\partial y'^2$, with $k_0=2\pi/\lambda_0$. 
We now translate this equation to ``physical space'' to obtain a transformed Helmholtz equation that features the same field distribution $E$ in the physical coordinates ($x$, $y$)
and in the inhomogeneous (but isotropic) refractive index landscape $n^2(x,y)$:
\begin{align}\label{eq:solphys}
\Delta E(x,y)+n^2(x,y)\,k_0^2\,E(x,y)=0,
\end{align}
where $\Delta=\nabla^2=\partial^2/\partial x^2+\partial^2/\partial y^2$. If we do not restrict ourselves to conformal (i.e., angle-preserving) transformations, the refractive index $n(x,y)=n_R(x,y)+in_I(x,y)$, which is now a complex-valued function in general, satisfies the following complicated relation that notably depends not only on the virtual index $n_0$ and the coordinate transformation [$x'(x,y)$, $y'(x,y)$], 
but also on the specific solution $E(x',y')$ in virtual space  
(see the Supplemental Material - SM for details):
\begin{align}\label{eq:refr_ind}
\begin{array}{l}
n^2(x,y)=\frac{n_0^2}{2}\left[(\nabla x')^2+(\nabla y')^2\right]-\frac{1}{k_0^2}\left(\Delta x'\frac{\partial \ln E}{\partial x'}+\Delta y'\frac{\partial \ln E}{\partial y'}\right) \\[.2cm]
-\frac{2}{k_0^2}\left(\nabla x'\cdot\nabla y'\right)\left( \frac{\partial^2\ln{E}}{\partial x'\partial y'}+\frac{\partial \ln{E}}{\partial x'}\frac{\partial \ln{E}}{\partial y'}\right)-\frac{1}{2k_0^2}\left[(\nabla x')^2-(\nabla y')^2\right]\\[.2cm]
\times\left[\frac{\partial^2\ln{E}}{\partial x'^2}+\left(\frac{\partial \ln{E}}{\partial x'}\right)^2 -\frac{\partial^2\ln{E}}{\partial y'^2}-\left(\frac{\partial \ln{E}}{\partial y'}\right)^2\right] .
\end{array}
\end{align}
Whereas the first term on the right-hand side of the above equation, $(n_0^2/2)[(\nabla x')^2+(\nabla y')^2]$, is the conformal part (see e.g. Ref.~\cite{leonhardt_geometry_2010}), the remaining terms stem from the non-conformality of the coordinate transformation. Since these new terms bind the transformation of the index to a certain solution in virtual space, $E(x',y')$, they appear very impractical at first glance. It turns out, however, that for the canonical case of a plane-wave input in positive $x$-direction, corresponding to a virtual solution $E(x',y')=E_0\,e^{in_0k_0x'}$, the optical potential in real space drastically simplifies to take the intensity-independent form:  
\begin{align}\label{eq:ci_pot}
n^2(x,y)=n_0^2\left[(\nabla x')^2-\frac{i}{n_0k_0}\Delta x'\right].
\end{align}
Remarkably, this expression for $n^2(x,y)$ is equivalent to the 2D constant-intensity (CI) potential that has recently been studied extensively \cite{makris_constant-intensity_2015,makris_wave_2017,yu_bohmian_2018,rivet_constant-pressure_2018,sebbah_scattering_2017,brandstotter_scattering-free_2019,horsley_indifferent_2019,makris_scattering-free_2020,tzortzakakis_shape-preserving_2020} (see the SM). CI waves are a special solution of the Helmholtz equation for which a judiciously chosen modulation of gain and loss suppresses wave scattering entirely. Moreover, as was recently shown in \cite{brandstotter_scattering-free_2019,makris_scattering-free_2020}, the media supporting such waves can be made uni-directionally invisible for a broad range of input frequencies. With Eq.~(\ref{eq:ci_pot}) we have thus discovered that CI-waves in physical space arise naturally through a non-conformal coordinate transformation of the plane wave solution in homogeneous virtual space. This provides a simple geometrical interpretation of CI-waves, and some of their hitherto unexplained properties, such as their robustness to frequency detuning, can be seen as a direct consequence of this interpretation. 

\begin{figure}[!h]
\centering
\includegraphics[clip,width=\columnwidth]{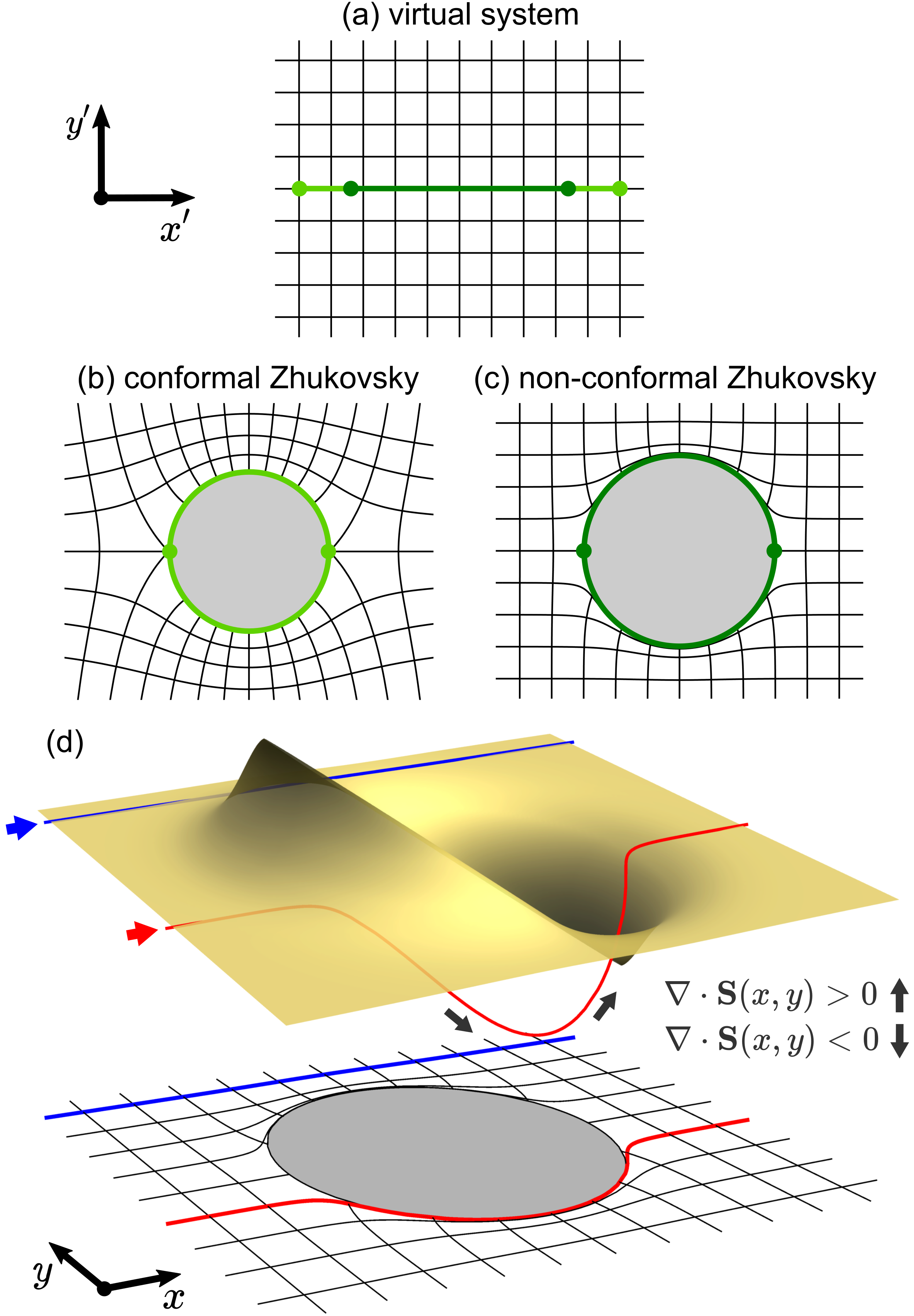}
\caption{Transforming space with non-Hermitian media. (a) The virtual coordinates, with the branch cuts of the conformal (light green) and non-conformal (dark green) Zhukovsky maps. The area of each square in the grid is $0.4 \times 0.4$. After the mappings, the green lines form circles, see (b) and (c), inside of which is the cloaked region (gray shaded area). The local orthogonality of the coordinate lines, a signature of conformal mapping, is present only in (b), but not in (c) (in both plots only the upper Riemann sheet is depicted). (d) Visualization of light propagation in the proposed non-Hermitian cloak profile, with the blue and red lines being the equivalent of rays in a non-Hermitian landscape (see the SM). The yellow surface indicates the imaginary part of the dielectric function, that causes the blue and red lines to rise or fall in the direction orthogonal to the $x-y$ plane, indicating the local creation ($\nabla\cdot\mathbf{S}(x,y)>0$) or destruction ($\nabla\cdot\mathbf{S}(x,y)<0$) of energy flux. The projection of the lines onto the 2D plane coincides with the inversely transformed coordinates, plotted in the grid below, with the cloaked region shaded in gray.}\label{Fig:1}
\end{figure}

In 2D virtual space not only plane waves, but also other continuous wave solutions can be identified, for which the refractive index of Eq.~(\ref{eq:refr_ind}) will be independent of the virtual beam's intensity. Such solutions typically have separable amplitude-dependent and amplitude-independent parts, another example of which is a Gaussian beam. On the other hand, electric field solutions with diverging logarithmic derivatives, such as Bessel \cite{durnin1987diffraction} and Airy beams \cite{siviloglou_observation_2007}, will create regions of infinite $n(x,y)$, and are hence unsuitable for our transformation protocol. 

\begin{figure}[!h]
\centering
\includegraphics[clip,width=\columnwidth]{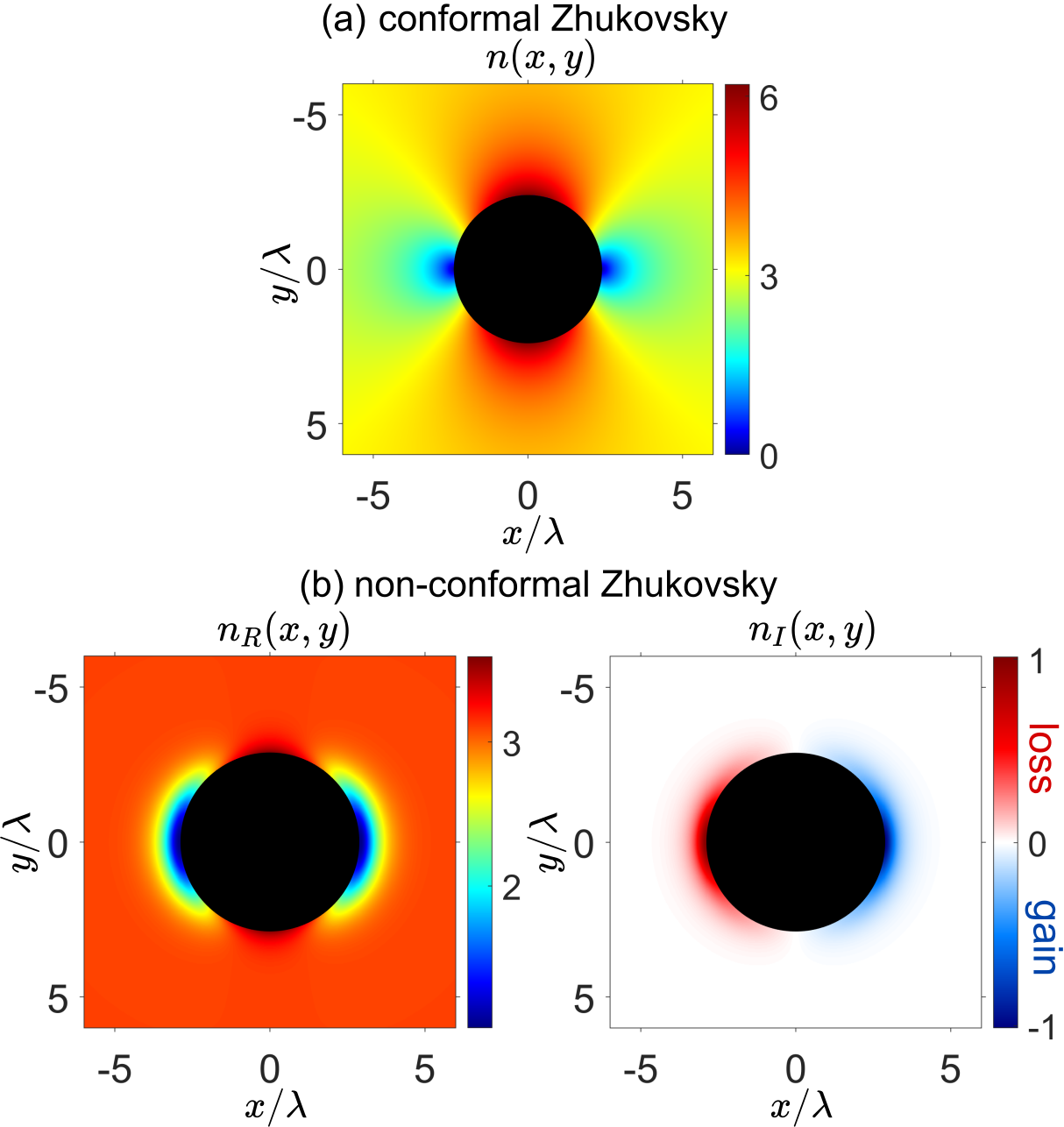}
\caption{Refractive index distributions for the conformal and non-conformal Zhukovsky cloaks. (a) Real-valued refractive index distribution of the conformal Zhukovsky cloak, reaching values of $n=0$ at $x=\pm 2.4\lambda$, and requiring the use of ENZ materials. Here, $\lambda=\lambda_0/n_0$ and $n_0=3.1$. (b) Real (left) and imaginary (right) parts of the refractive index spatial profile of the non-conformal Zhukovsky map. The refractive index based on Eq.~(\ref{eq:ci_pot}) is calculated using the transformation $w(z,z^*)$ with $\eta(z,z^*) = 1/[1+e^{\beta(|z|-R_1)}]-1/[1+e^{\beta(|z|-R_0)}]$ and $R_0=0.25$, $R_1=1$, $\beta=5.75$. The values for $n_R(x,y)$ lie between $1.013$ and $3.594$, whereas those of $n_I(x,y)$ lie between $-1.013$ and $1.013$. The black filled circle in all subfigures indicates the cloaked region.}\label{Fig:2}
\end{figure}


\begin{figure}[t!]
\centering
\includegraphics[clip,width=\columnwidth]{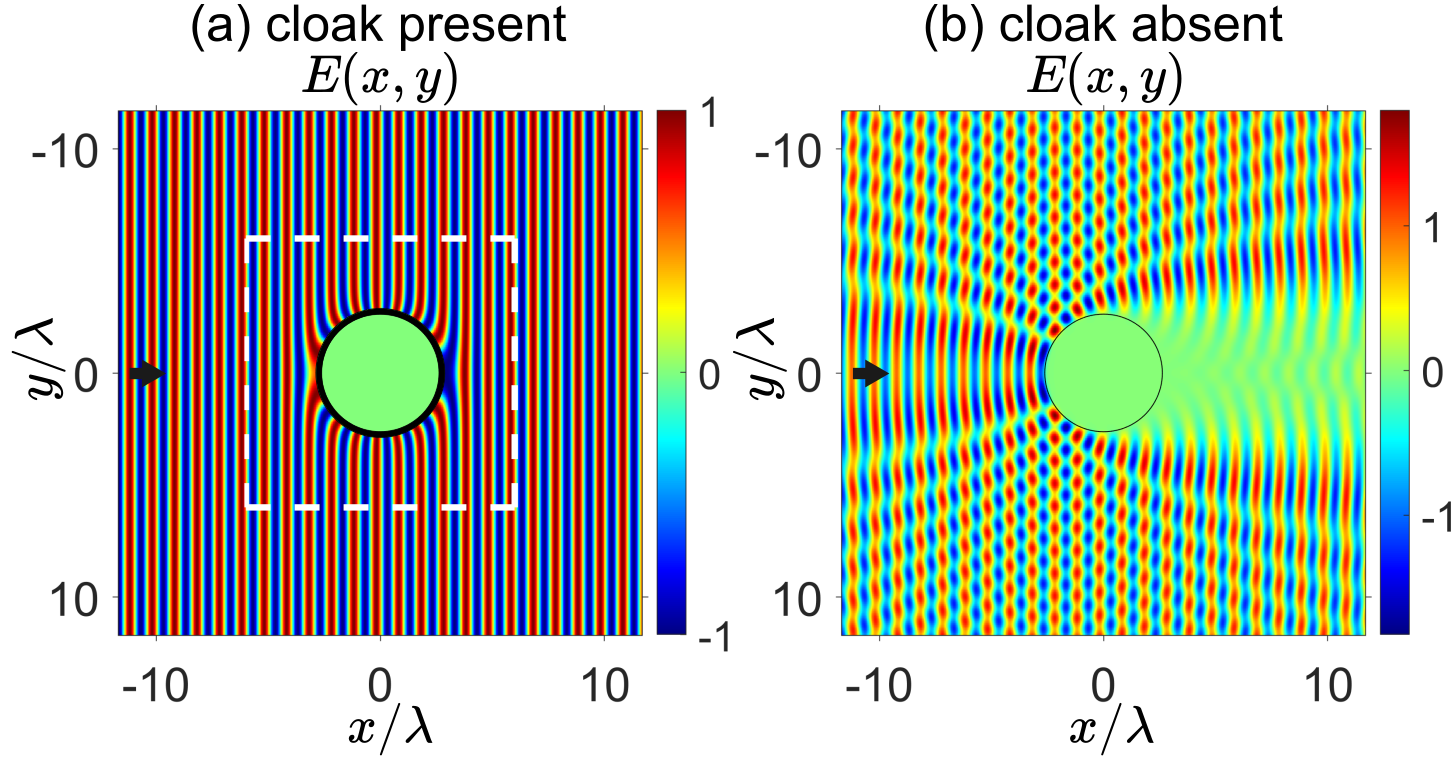}
\caption{Non-Hermitian cloaking for an input plane wave at the center design spatial frequency $k=n_0k_0$. (a) Electric field solution for the case when cloaking is present. The refractive index distribution is that of  Fig.~\ref{Fig:2}(b), with the dashed white box denoting the area plotted there. The black ring represents the annulus of the cloak, where the Neumann boundary condition of vanishing normal electric field derivative was used. Inside the annulus there is a highly reflective material, taken here as aluminium ($n_{Al}=1.52+9.26i$ at $1\mu m$ \cite{rakic1998optical}), with a circular cross section. (b) The electric field solution for the same input field but with only the  highly reflective material present. The cloak's outer radius is $2.86\lambda$, while its inner radius is $2.62\lambda$. The fields are normalized to the input wave amplitude. The black arrow marks the propagation direction of the input beam. }\label{Fig:3}
\end{figure} 
We now apply the above approach to a well-studied example in the literature on transformation optics, which is the Zhukovsky map \cite{zhukovsky_uber_1912} that has been used extensively in conformal transformation optics for the design of invisibility cloaks \cite{leonhardt_optical_2006,leonhardt_geometry_2010,xu_conformal_2015}. We start with the expression for the conformal Zhukovsky map, $w_{zh}(z)=z+1/z$, where we have used the convenient notation $w=x'+iy'$ for the virtual coordinates, and $z=x+iy$ for the physical coordinates [note that ($x',y'$) and ($x,y$) are real-valued, see Refs.~\cite{castaldi_p_2013,horsley_wave_2016,savoia2016} for examples of complex-valued transformations]. Illustrations of the conformal Zhukovsky transformation are provided in Figs.~\ref{Fig:1}(a,b): the light green line in virtual space, see Fig.~\ref{Fig:1}(a) (branch cut connecting the points $w=\pm 2$), is transformed into the light green unit circle in physical space, see Fig.~\ref{Fig:1}(b). A plane wave traveling in positive $x$-direction of virtual space will thus be guided around the cloak boundary, placed right at this circle in physical space. Importantly for the present case, the conformal Zhukovsky map features two points located at $z=\pm 1$, right at the cloak boundary, where the refractive index in physical space vanishes [see light green points in Fig.~\ref{Fig:1}(b) and dark blue parts in Fig.~\ref{Fig:2}(a)]. The cloak thus requires the use of materials with a vanishing refractive index, which limits the cloak functionality to narrow-band radiation.

With the approach presented above, such limitations can be conveniently circumvented through the use of non-conformal maps that naturally lead to non-Hermitian materials with complex refractive profiles. In particular, to design such a non-Hermitian cloak, we modify the Zhukovsky map as $w(z,z^*) = z +\eta(z,z^*)/z$.

In contrast to the conformal map $w_{zh}(z)$, the map $w(z,z^*)$ depends on both $z$ and $z^*$, which violates conformality (compare Figs.~\ref{Fig:1}(b) and~\ref{Fig:1}(c)). In the present case, we choose the envelope $\eta(z,z^*)$ as a function of $|z|$ to have the shape of a flat-top function with smoothed edges (see caption of Fig.~\ref{Fig:2}). For large values of $|z|\gg 1$ the envelope $\eta(z,z^*)$ vanishes  [$w(z,z^*)\to z$], while at $|z|\approx 1$ it has an edge with finite first and second derivatives, such that the vanishing of $n_R$ at $x=\pm 1=\pm 2.4\lambda$), shown for the conformal Zhukovsky map in Fig. \ref{Fig:2}(a), is avoided, according to Eq.~(\ref{eq:ci_pot}). We have found that for a judicious choice of $\eta(z,z^*)$ excellent cloaking can be achieved when the cloak's outer radius corresponds to the radius of the branch cut in physical space. For the cloak we show here, this radius is $|z|=1.20458=2.86\lambda$, and is indicated by the dark green line and circle in Fig.~\ref{Fig:1}(a) and (c), respectively. The resulting material is dielectric with $n_R(x,y)={\mathrm{Re}}[n(x,y)]>1$ in all points of physical space, but with an inhomogeneous gain/loss distribution, given by $n_I(x,y)={\mathrm{Im}}[n(x,y)]$. 

The principle of the cloak's operation is schematically depicted in Fig.~\ref{Fig:1}(d). The yellow surface represents the imaginary part of the dielectric profile,  $\varepsilon_I(x,y)=\mbox{Im}[n^2(x,y)]$. As the energy flux is created and destroyed in regions of gain and loss, the representation of light as rays in 2D space is not appropriate in non-Hermitian media. To visualize the light propagation, we plot instead the lines whose increase (decrease) in the direction orthogonal to the $x\!-\!y$ plane marks the creation (destruction) of the Poynting flux $\mathbf{S}(x,y)$ (see the SM). The cloaking for a beam incoming from the negative $x$ direction [red line in the upper plot of Fig. \ref{Fig:1}(d)] can be explained by the interplay of the real and imaginary parts of the refractive index distribution, which form here a parity-time (PT) symmetric system [see Fig.~\ref{Fig:2}(b)]. The part of the beam that is sufficiently displaced from the cloak center (blue line) sees a non-curved space, and propagates in a straight line in the homogeneous background medium. For beam parts near the center (red line), similarly to the conformal Zhukovsky mapping, the real part of the refractive index distribution works to bend the light around the object. Since, however, we have $n_R>1$ in all of physical space, the real part of the refractive index alone is insufficient to achieve the cloaking effect: in fact, the finite imaginary part of the refractive index distribution partially absorbs the incoming electric field [$\nabla\cdot\mathbf{S}(x,y)<0$] in front of the cloak and amplifies it [$\nabla\cdot\mathbf{S}(x,y)>0$] behind the cloak. 
\begin{figure}[t!]
\centering
\includegraphics[clip,width=\columnwidth]{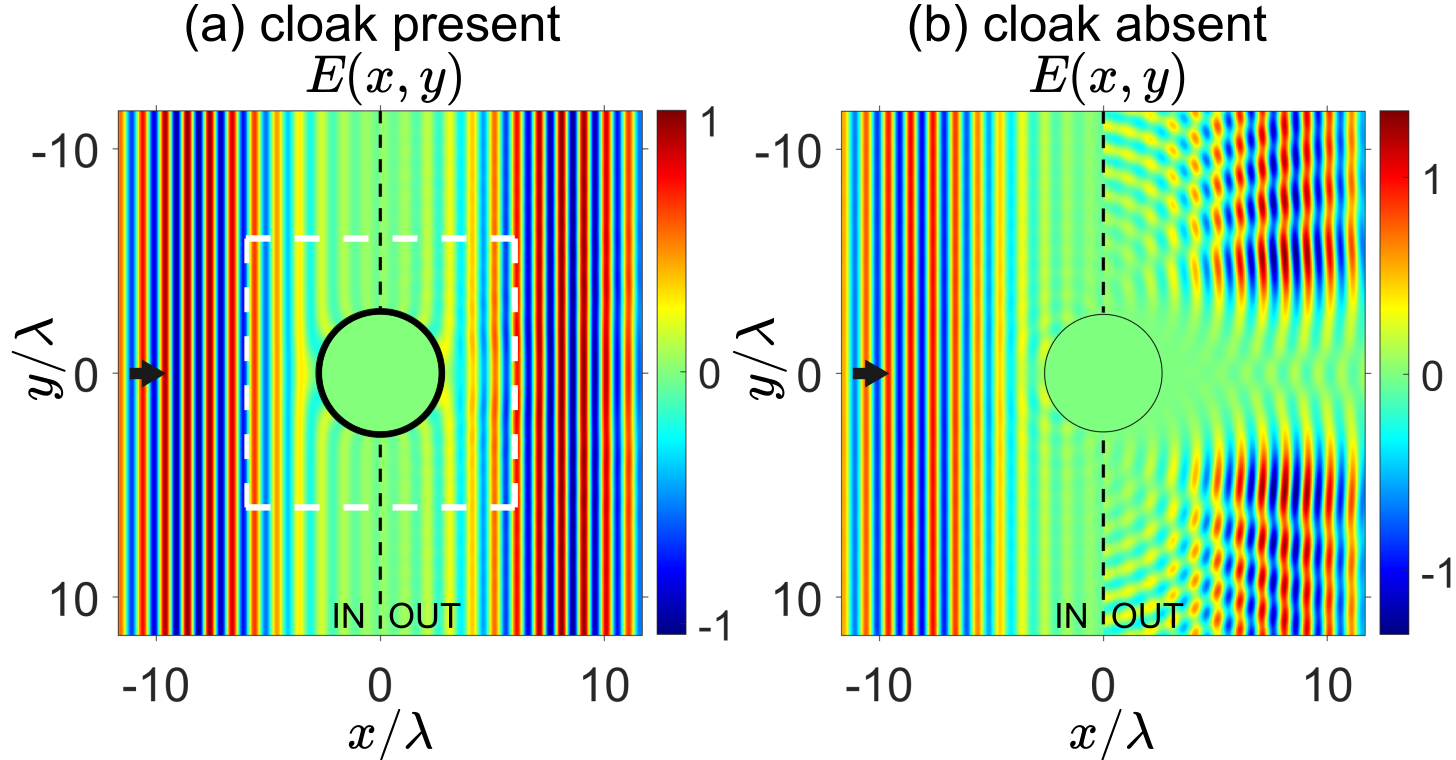}
\caption{Demonstration of cloaking under pulsed illumination, for the optical media as in Fig.~\ref{Fig:3}. (a) A spatially narrow pulse incident onto the cloaking region (left) is nearly perfectly transmitted (right) when the cloak is present. (b) When the cloak is absent, there is a distortion of the pulse shape, clearly revealing the presence of the reflecting material ($n_{Al}=1.52+9.26i$) to the outside observer. The thin black dashed line denotes the separation of the plotting regions for the incoming (left) and outgoing (right) part of the pulse evolution, all other markings are as in Fig.~\ref{Fig:3}. The time evolution was calculated by an inverse Fourier transform, using a Gaussian envelope with a width of $\sigma_k=0.055k$ in $k$-space.
}\label{Fig:4}
\end{figure}

To demonstrate the cloaking efficiency, we solve  Eq.~(\ref{eq:solphys}) with $n(x,y)$ of Fig.~\ref{Fig:2}(b) numerically by using the open source finite-element solver NGSolve \cite{schoberl_netgen_1997,schoberl2014}.  Inside of the cloaked region that we terminate with a Neumann boundary condition in physical space, we place a highly reflective material with a circular cross-section. Choosing the background value of the refractive index in virtual space as $n_0=3.1$, we find that $n_R(x,y)$ varies between $1.013$ and $3.594$, whereas $n_I(x,y)$ has values between $\pm 1.013$. In the SM, we also show the results for an alternative parameter regime, where the cloak region has a smaller radius but the $n_I(x,y)$ varies with an amplitude as low as $\pm 0.08$. 

When the cloak is present [Fig.~\ref{Fig:3}(a)], the beam incoming from the left is perfectly transmitted to the right side, with both phase and amplitude in the far field being equal to that of a plane wave in virtual space with a refractive index $n_0$ [see also Fig.~S2 of the SM)]. Inside the cloaked area, the field completely vanishes, as expected. When the cloak is absent [Fig.~\ref{Fig:3}(b)], strong scattering occurs and a shadow is formed behind the cloak. 

Concerning the directional sensitivity of the cloak operation, we recall that already the conformal Zhukovsky cloak is sensitive to changes of the input angle with respect to the normal axis (see the SM). The cloak in Fig.~\ref{Fig:2}(b) retains a similar sensitivity to such deviations, with an appreciable reduction of efficiency already at input beam tilts of around $\pm2^\circ$ to the normal, making the cloaking effectively unidirectional. 

To put these results in the context of previous theoretical work, let us mention here Ref.~\cite{zhu_one-way_2013}, where a cloak based on anisotropic non-Hermitian electric and magnetic materials was designed by transforming a PT-symmetric potential in the virtual coordinates to the physical space—an idea later transferred also to acoustics~\cite{zhu2014p}. In a similar vein, folding and stretching of non-Hermitian virtual space has recently been used to construct 2D gain-loss distributions that are robustly balanced \cite{luo2021}. In Ref.~\cite{sounas_unidirectional_2015}, a cloaking strategy based on subwavelength layers of balanced gain and loss was devised, using conducting non-Hermitian metamaterials with locally infinite reflection coefficients. This strategy, although similar in spirit to active cloaking, does not require the knowledge of the input wave properties for successful cloak operation, in contrast to the earlier work of Refs.~\cite{miller2006,selvanayagam2013}. We note here that none of the earlier non-Hermitian cloaking work, that we are aware of, involves transformation optics with isotropic dielectric media, as we do in this Letter. First remarkable successes in the direction of non-Hermitian transformation optics were made in one-dimensional (1D) systems, using complex spatial coordinates  \cite{castaldi_p_2013,savoia2016,horsley_wave_2016}.

A notable disadvantage afflicting ENZ~\cite{leonhardt_optical_2006}, anisotropic~\cite{schurig_metamaterial_2006} and layered non-Hermitian cloaks \cite{sounas_unidirectional_2015} alike, is the requirement of extreme real refractive index values (or admittance values \cite{sounas_unidirectional_2015}), which limits the cloak's operation to frequencies near metamaterial resonances. As our cloak features inhomogeneous dielectric media with $n_R>1$ and finite values of $n_I$, which are known to have reasonably fast and frequency-broadband responses (see, e.g., Refs.~\cite{moloney2007quantum,guzelturk2018giant}), we now investigate the behavior of our cloak under pulsed illumination. 

Strictly speaking, the cloak's refractive index distribution in Eq.~(\ref{eq:ci_pot}) should produce its desired effect only at the design wavenumber $k=n_0k_0$. Perfect cloaking from pulsed radiation would thus require the tailoring of not only the spatial but also the frequency dependence of the refractive index distribution. We choose here to approximate the spectral index distribution to be entirely frequency-independent in the $k$-range of the input pulse, while using the spatial distribution shown in Fig.~\ref{Fig:2}(b). The results, shown in Fig.~\ref{Fig:4}(a), demonstrate excellent cloaking performance for pulses with a spectral width of $\sigma_k=0.055\, k$, thereby confirming the robustness of the design in Eq.~(\ref{eq:ci_pot}) to frequency detuning. 

To summarize, we have introduced a new way to achieve optical cloaking in dielectric systems without using ENZ or anisotropic materials, and demonstrated that non-conformal maps naturally lead to complex isotropic indices of refraction. Based on a non-conformal Zhukovsky transformation, a non-Hermitian invisibility cloak is introduced, which also works for incoming pulses. We note that our transformation optics design is fully analytical, but still optimization strategies could also be applied in order to optimize the corresponding refractive index profiles, with relatively low computational effort as compared to previous work in non-Hermitian optimization problems \cite{bachelard2012,hisch2013,bachelard2014}. Although the discussed synergy between non-Hermitian photonics and 2D isotropic transformation optics is expected to lead to many more interesting insights, we emphasize here that it should be possible to extend our theoretical methodology to three spatial dimensions, where the vectorial properties of light-matter interaction play a significant role.

\begin{acknowledgments}
The work of I.K.\ was funded by the Austrian Science Fund (FWF) Lise Meitner Postdoctoral Fellowship M3011 and the FWF grant P32300. The joint work of I.K., K.G.M. and S.R. was supported by the European Commission grant MSCA-RISE 691209. U.L.\ was supported by the Israel Science Foundation and the Murray B.\ Koffler Professorial Chair. The computational results presented here have been obtained using the Vienna Scientific Cluster (VSC).
\end{acknowledgments}

\clearpage
\onecolumngrid

\renewcommand{\thefigure}{S\arabic{figure}}
\renewcommand{\theequation}{S\arabic{equation}}

\setcounter{equation}{0}
\setcounter{figure}{0}
\section{Supporting information: Transforming space with non-Hermitian dielectrics}

\section{Non-conformal transformation optics in isotropic media}
Here we provide a detailed derivation of the Eq.~(3) of the main text, which is related to the geometric properties of the transformed space. In particular, the Helmholtz equation in our physical system, which is an isotropic medium described by a real- or complex-valued refractive index profile $n(x,y)$, is the following (using normalized units):
\begin{align}\label{eq:helm_physinit}
\frac{ \partial^2 E}{\partial x^2}+\frac{ \partial^2 E}{\partial y^2}+n^2\,k_0^2\,E=0,
\end{align}
where $E(x,y)$ is the electric field in a two-dimensional space. Such an equation can be written in a more compact form, by using complex variables as:
\begin{align}\label{eq:helm_phys}
4\frac{ \partial^2}{\partial z\partial z^*}E+n^2\,k_0^2\,E=0,
\end{align}
by writing $z=x+iy$ and using the definition of Wirtinger derivatives:
\begin{align}\label{eq:coords_phys}
\frac{ \partial}{\partial z}=\frac{1}{2}\left(\frac{\partial}{\partial x}-i\frac{\partial}{\partial y}\right),\:\:\quad \frac{ \partial}{\partial z^*}=\frac{1}{2}\left(\frac{\partial}{\partial x}+i\frac{\partial}{\partial y}\right).
\end{align}
Moving now to the primed (virtual) coordinates $w=x'+iy'$, the Hemholtz equation is given by:
\begin{align}\label{eq:helm_virt}
4\frac{ \partial^2}{\partial w\partial w^*}E+n'^2\,k_0^2\,E=0.
\end{align}
With the coordinate transformation $w(z,z^*)=x'(x,y)+iy'(x,y)$, based on the identities
\begin{align}
\frac{ \partial}{\partial z}=\frac{\partial w}{\partial z}\frac{\partial }{\partial w}+\frac{\partial w^*}{\partial z}\frac{\partial }{\partial w^*},\:\:\quad \frac{ \partial}{\partial z^*}=\frac{\partial w}{\partial z^*}\frac{\partial }{\partial w}+\frac{\partial w^*}{\partial z^*}\frac{\partial }{\partial w^*},
\end{align}
the two-dimensional Laplacian is transformed as
\begin{align}\label{eq:laplacian}
\frac{1}{4}\Delta&=\frac{\partial^2}{\partial z\partial z^*}=\frac{\partial w}{\partial z}\frac{\partial w}{\partial z^*}\frac{\partial ^2}{\partial w^2}+\left( \left|\frac{\partial w}{\partial z} \right|^2+\frac{\partial w^*}{\partial z}\frac{\partial w}{\partial z^*}\right)\frac{\partial^2}{\partial w\partial w^*}
+\frac{\partial w^*}{\partial z}\frac{\partial w^*}{\partial z^*}\frac{\partial ^2}{\partial w^{*2}}+\frac{\partial^2 w}{\partial z\partial z^*}\frac{\partial}{\partial w}+\frac{\partial^2 w^*}{\partial z\partial z^*}\frac{\partial}{\partial w^*}
\,.
\end{align}
By choosing $w$ not to depend on $z^*$ (and hence also $w^*$ not to depend on $z$), we are left with $\Delta=\left|\frac{\partial w}{\partial z} \right|^2\Delta '$, and the refractive indices in the two media are related by a spatially-varying factor for all 2D waves at $k_0$, giving $n=\left|\frac{\partial w}{\partial z} \right| n'$ \cite{leonhardt_geometry_2010}. It can be shown that the Cauchy-Riemann conditions are satisfied in this case, so that the coordinate transformation $(x,y)\to (x',y')$ is conformal, and the propagation satisfies the Fermat principle of shortest optical path in an isotropic medium.

When using a non-Hermitian medium with finite gain and loss, the above simplification, stemming from the Fermat principle in isotropic media, may not apply, and thus the simple relation between Laplacians in the two coordinate systems is not satisfied. However, one can still write the Helmholtz equation for a particular solution $E$, as is described in details below.

To show this, we write the derivatives in the above equation in the transformed coordinates as
\begin{align}
\begin{array}{c}
\frac{ \partial^2}{\partial w^2}=\frac{1}{4}\left( \frac{\partial^2}{\partial x'^2}-  \frac{\partial^2}{\partial y'^2} -2i\frac{\partial^2}{\partial x'\partial y'}\right),\\[0.2cm]
\frac{\partial^2}{\partial w\partial w^*}=\frac{1}{4}\Delta ',\\[0.2cm]
\frac{ \partial^2}{\partial w^{\star 2}}=\frac{1}{4}\left( \frac{\partial^2}{\partial x'^2}-  \frac{\partial^2}{\partial y'^2} +2i\frac{\partial^2}{\partial x'\partial y'}\right).
\end{array}
\end{align}
Writing the derivatives of the electric field $E$ as
\begin{align}
\begin{array}{c}
\frac{ \partial^2E}{\partial x'^2}=\left[ \frac{\partial^2\ln{E}}{\partial x'^2}+\left(\frac{\partial \ln{E}}{\partial x'}\right)^2 \right]E,\\[0.3cm]
\frac{ \partial^2E}{\partial x'\partial y'}=\left[ \frac{\partial^2\ln{E}}{\partial x'\partial y'}+\frac{\partial \ln{E}}{\partial x'}\frac{\partial \ln{E}}{\partial y'} \right]E,\\ [0.3cm]
\frac{ \partial^2E}{\partial y'^2}=\left[ \frac{\partial^2\ln{E}}{\partial y'^2}+\left(\frac{\partial \ln{E}}{\partial y'}\right)^2 \right]E,
\end{array}
\end{align}
the transformation of the Laplacian (\ref{eq:laplacian}) allows us to transform  Eq.~(\ref{eq:helm_phys}) to the Helmhotz equation (\ref{eq:helm_virt}) of the virtual system, with a new refractive index $n'$ related to $n$ as:
\begin{align}
\begin{array}{c}
n'^{\,2}=\left(\left|\frac{\partial w}{\partial z}\right|^2+\frac{\partial w^*}{\partial z}\frac{\partial w}{\partial z^*}\right)^{-1}\left[ n^2+ \frac{1}{k_0^2}\left(\frac{\partial w}{\partial z}\frac{\partial w}{\partial z^*} +\frac{\partial w^*}{\partial z}\frac{\partial w^*}{\partial z^*}\right)\right.
\times\left(\frac{\partial^2\ln{E}}{\partial x'^2}+\left(\frac{\partial \ln{E}}{\partial x'}\right)^2 -\frac{\partial^2\ln{E}}{\partial y'^2}-\left(\frac{\partial \ln{E}}{\partial y'}\right)^2\right) \\[0.2cm]
+\frac{2i}{k_0^2}\left( \frac{\partial^2\ln{E}}{\partial x'\partial y'}+\frac{\partial \ln{E}}{\partial x'}\frac{\partial \ln{E}}{\partial y'} \right)\left( \frac{\partial w^*}{\partial z}\frac{\partial w^*}{\partial z^*}-\frac{\partial w}{\partial z}\frac{\partial w}{\partial z^*}  \right) 
+\frac{2}{k_0^2}\left.\left(\frac{\partial^2(w+w^*)}{\partial z\partial z^*}\frac{\partial\ln E}{\partial x'}-i\frac{\partial^2(w-w^*)}{\partial z\partial z^*}\frac{\partial\ln E}{\partial y'}\right)\right].
\end{array}
\end{align}
This relation can be written as:
\begin{align}
\begin{array}{c}
n'^{\,2}=\frac{2}{(\nabla x')^2+(\nabla y')^2}\left\{ n^2+ \frac{1}{k_0^2}\left(\Delta x'\frac{\partial \ln E}{\partial x'}+\Delta y'\frac{\partial \ln E}{\partial y'}\right)+\frac{1}{2k_0^2}\left[(\nabla x')^2-(\nabla y')^2\right]\right.
\left[\frac{\partial^2\ln{E}}{\partial x'^2}+\left(\frac{\partial \ln{E}}{\partial x'}\right)^2 -\frac{\partial^2\ln{E}}{\partial y'^2}-\left(\frac{\partial \ln{E}}{\partial y'}\right)^2\right]\\[0.2cm]
+\frac{2}{k_0^2}\left.\left( \nabla x'\cdot\nabla y'\right)\left( \frac{\partial^2\ln{E}}{\partial x'\partial y'}+\frac{\partial \ln{E}}{\partial x'}\frac{\partial \ln{E}}{\partial y'} \right)\right\}.
\end{array}
\end{align}
If, as in the main text and the rest of the Supplemental Material, the virtual medium i homogeneous with $n'=n_0$, the above equation simplifies to the Eq. (3) of the main text:
\begin{align}
\begin{array}{c}
n^2(x,y,E)=\frac{n_0^2}{2}\left[(\nabla x')^2+(\nabla y')^2\right]-\frac{1}{k_0^2}\left(\Delta x'\frac{\partial \ln E}{\partial x'}+\Delta y'\frac{\partial \ln E}{\partial y'}\right)\\[0.2cm]
-\frac{1}{2k_0^2}\left[(\nabla x')^2-(\nabla y')^2\right]
\left[\frac{\partial^2\ln{E}}{\partial x'^2}+\left(\frac{\partial \ln{E}}{\partial x'}\right)^2 -\frac{\partial^2\ln{E}}{\partial y'^2}-\left(\frac{\partial \ln{E}}{\partial y'}\right)^2\right]
 -\frac{2}{k_0^2}\left(\nabla x'\cdot\nabla y'\right)\left( \frac{\partial^2\ln{E}}{\partial x'\partial y'}+\frac{\partial \ln{E}}{\partial x'}\frac{\partial \ln{E}}{\partial y'}\right).
\end{array}
\end{align}

\section{Relationship to constant-intensity waves}
In the Supplementary Material of Ref.~\cite{makris_scattering-free_2020} we have shown that a 2D constant-intensity wave solution $E_{CI}=Ae^{ik_0\theta_{CI}(x,y)}$ exists in an inhomogeneous non-Hermitian medium described by a dielectric function $\varepsilon_{CI}(x,y)=n_{CI}^2(x,y)$:
\begin{align}\label{eq:CI_diel}
n_{CI}^2(x,y)=(\nabla \theta_{CI})^2-\frac{i}{k_0}\Delta\theta_{CI}.
\end{align}
This corresponds to Eq.~(4) of the main text:
\begin{align}\label{eq:CI}
n^2(x,y)=n_0^2\left[(\nabla x')^2-\frac{i}{n_0k_0}\Delta x'\right],
\end{align}
with the following relationship between the phase profile and the coordinate transformation:  $\theta_{CI}(x,y)=n_0x'(x,y)$.

\section{Visualizing beam propagation for non-conformal Zhukovsky transformation}
Although the ray trajectories can only be defined for Hermitian media, as mentioned in the main text, here we show that even in non-Hermitian media one can still relate the geometric properties of the optical space to the shape of the electric field solution $E=E_ze^{i\varphi}$ of the Helmholtz equation. We take $\mathbf{r}_h(t)=[x_h(t),y_h(t)]$ to be a curve in the 2D physical space which is asymptotically a horizontal line, and $\mathbf{r}_v(t)=[x_v(t),y_v(t)]$ to be a curve in the same space which is asymptotically a vertical line, where $t$ is a parameter. These curves are defined as lines to which the local wave vector is tangent to throughout all space, which in Hermitian media means they are rays (see e.g. Chapter 30 of Ref.~\cite{leonhardt_geometry_2010}). In isotropic materials, the curves follow Hamilton's equations with $\omega=c|\mathbf{k}|=c\sqrt{k_x^2+k_y^2}$. Here, $\mathbf{k}=\nabla\varphi$, which for $\varphi=n_0k_0x'(
x,y)-\omega T$, where $T$ is a time variable, gives 
\begin{align}\label{eq:rhor}
\frac{d\mathbf{r}_h}{dt}=\frac{d}{dt}\begin{pmatrix} x_h\\[.1cm]
y_h\end{pmatrix} =\begin{pmatrix}\frac{\partial x'}{\partial x}\\[.1cm]
\frac{\partial x'}{\partial y}\end{pmatrix},
\end{align}
and for $\varphi=n_0k_0y'(
x,y)-\omega T$, it gives
\begin{align}\label{eq:rhor2}
\frac{d\mathbf{r}_v}{dt}=\frac{d}{dt}\begin{pmatrix} x_v\\[.1cm]
y_v\end{pmatrix} =\begin{pmatrix}\frac{\partial y'}{\partial x}\\[.1cm]
\frac{\partial y'}{\partial y}\end{pmatrix},
\end{align}
where we take the speed of light in the background medium equal to unity. These relations are valid for both conformal and non-conformal transformations. Although in Hermitian media these lines correspond to ray trajectories, this will not be the case for the non-Hermitian transformation media, as shown below. 

For the non-conformal Zhukovsky cloak, it can be shown that the Poynting vector is given by (see the Supplementary Material of Ref. \cite{makris_scattering-free_2020}):
\begin{align}
\mathbf{S}(x,y)=\frac{i}{2}(E \nabla E^{*}-E^{*}\nabla E)=n_0\,k_0\,\nabla x'. 
\end{align}
The horizontal lines are then defined by $\dot{\mathbf{r}}_h=\mathbf{S}/n_0k_0$. As the ray picture is a good description only in the Hermitian case, where the Poynting theorem gives $\nabla\cdot\mathbf{S}(x,y)=0$, the coordinate lines are no longer parallel to rays in non-Hermitian media, as rays are difficult to define in this case. Avoiding the complications of writing the Poynting theorem for non-Hermitian media, we have chosen to use another route for visualizing the beam propagation in the physical system. Instead of two dimensions, we instead add a third dimension $\xi$, such that the Poynting vector is now divergence free, satisfying   
\begin{align}
\nabla\cdot\mathbf{S}(x,y,\xi)=\frac{\partial S_x}{\partial x}+\frac{\partial S_y}{\partial y}+\frac{\partial S_\xi}{\partial \xi}=0. 
\end{align}
It can readily be shown that this is satisfied for $S_\xi=-n_0k_0\xi\Delta x'+f(x,y)$. Choosing $f(x,y)=0$, the propagation of these new ``rays'', incoming from $x=-\infty$, is then governed by
\begin{align}\label{eq:rhor}
\frac{d\mathbf{r}}{dt}=\frac{d}{dt}\begin{pmatrix} x\\[.1cm]
y \\[.1cm]
\xi\end{pmatrix} =\begin{pmatrix}\frac{\partial x'}{\partial x}\\[.1cm]
\frac{\partial x'}{\partial y}\\[.1cm]
-\xi\Delta x'\end{pmatrix}.
\end{align}
The excursion of these lines from the $x-y$ plane thus indicates the presence of non-Hermiticity in the system, as $\Delta x'(x,y)=-\frac{k_0}{n_0}\varepsilon_I(x,y)=(n_0k_0)^{-1}\nabla\cdot\mathbf{S}(x,y)$. It can easily be shown that $\xi(t)$ is given by: 
\begin{align}
\xi(t)=\exp\left({-\frac{1}{n_0k_0}\int_0^tdt'\nabla\cdot\mathbf{S}[x(t'),y(t')]}\right)\,,
\end{align}
where we have chosen the initial condition $\xi(0)=1$. In Fig.~1(d) of the main text, we plot the quantity $\delta \xi(t)=0.025[1-\xi(t)]$. In this case, an increase (decrease) in $\delta\xi$ indicates a positive (negative) 2D divergence of the Poynting vector, since:   
\begin{align}
\frac{d\delta\xi(t)}{dt}=\frac{0.025}{n_0k_0}\,\nabla\cdot\mathbf{S}[x(t),y(t)]\cdot\exp\left({-\frac{1}{n_0k_0}\int_0^tdt'\nabla\cdot\mathbf{S}[x(t'),y(t')]}\right).
\end{align}

\begin{figure}[t!]
\centering
\includegraphics[clip,width=\columnwidth]{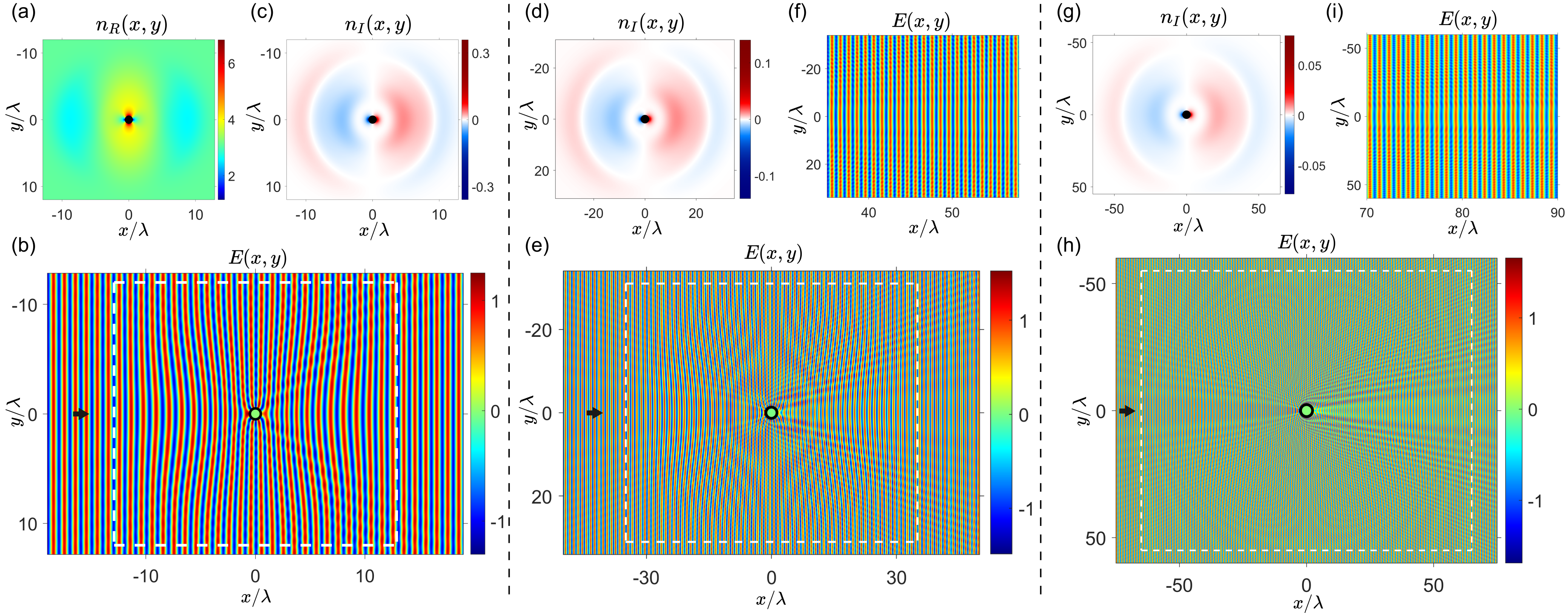}
\caption{Non-Hermitian cloaking for the same prescription as in the main text, but for parameter values: $R_0=1.5$, $R_1=4$, $\beta=2.0$. These parameter values produce a more complicated refractive index distribution, with a more gradual spatial variation, resulting in a thick cloak with smaller gain and loss modulation amplitude. The center operating wavelengths are here: (a)-(c) equal to the one in the main text, reduced by a factor of (d)-(f) 2.6667 and (g)-(i) 4.7273. The imaginary part of the refractive index distribution reaches values between (c) $[-0.36,0.36]$, (d) $[-0.14,0.14]$ and (g) $[-0.08,0.08]$. The electric field distribution for the factor of (b) 1, (e),(f) 2.6667 and (h),(i) 4.7273 (where (f) and (i) are the corresponding magnified views of the transmitted part of the wave) shows that the cloak retains its good efficiency when reducing the operating wavelength, and hence the gain and loss modulation amplitude. The black ring in (b), (e) and (h) marks the cloak layer, the black arrow marks the beam incidence direction,  while the white dashed box marks the regions plotted in the respective refractive index plots. The $n_R(x,y)$ modulation has the same shape as in (a) for all of the three operating wavelengths.}\label{Fig:regime2}
\end{figure} 

\section{Reducing the gain/loss modulation amplitude with a thick cloak}
In addition to the thin cloak design, an example of which is presented in the main text, we here consider another regime of cloaking, for which the refractive distribution occupies a larger area (see Fig.~\ref{Fig:regime2}). As a result of this more gradual non-conformal coordinate transformation, which has the same form as the transformation in the main text but with different parameter values, the gain/loss modulation is weaker in this case than for the thin cloak. The results in Fig.~\ref{Fig:regime2}(b) indicate that this cloak also works well in concealing the scattering object. The small modulation of the wave amplitude around the cloaking region is here a consequence of the fact that the radius of the cloak is slightly larger than the radius of the branch cut. This larger radius was selected in order to reduce the gain/loss modulation amplitude even further. 

An advantage of this type of non-conformal Zhukovsky cloak stems from the fact that the  dielectric function ($\varepsilon_I$), given by the imaginary part of Eq.~(\ref{eq:CI}), is nearly an order of magnitude smaller than the real part ($\varepsilon_R$) in all of space. One can thus apply the following approximation:
\begin{align}
n(x,y)=n_0|\nabla x'|\sqrt{\left[1-\frac{i}{n_0k_0}\frac{\Delta x'}{(\nabla x')^2}\right]}\approx n_0\left(|\nabla x'|-\frac{i\lambda}{4\pi}\frac{\Delta x'}{|\nabla x'|}\right).
\end{align}
This interesting property of our design strategy implies that the magnitude of the $n_I$ spatial modulation will fall almost linearly with reducing the design wavelength, for a chosen non-conformal transformation [$x'(x,y)$, $y'(x,y)$]. 

In Fig.~\ref{Fig:regime2}(d)-(i) we show how the cloak changes when the wavelength is reduced by a factor of 2.6667 and 4.7273. As can be seen, the efficiency of the cloaking is nearly the same but for a gain/loss modulation that is significantly reduced compared to the values in the main text. This property could prove crucial for an experimental implementation of our cloaking strategy. An optimization of the non-conformal mapping along these lines could enable a cloak design with a refractive index having a realistic range of values at optical frequencies.

\begin{figure}[t!]
\centering
\includegraphics[clip,width=\columnwidth]{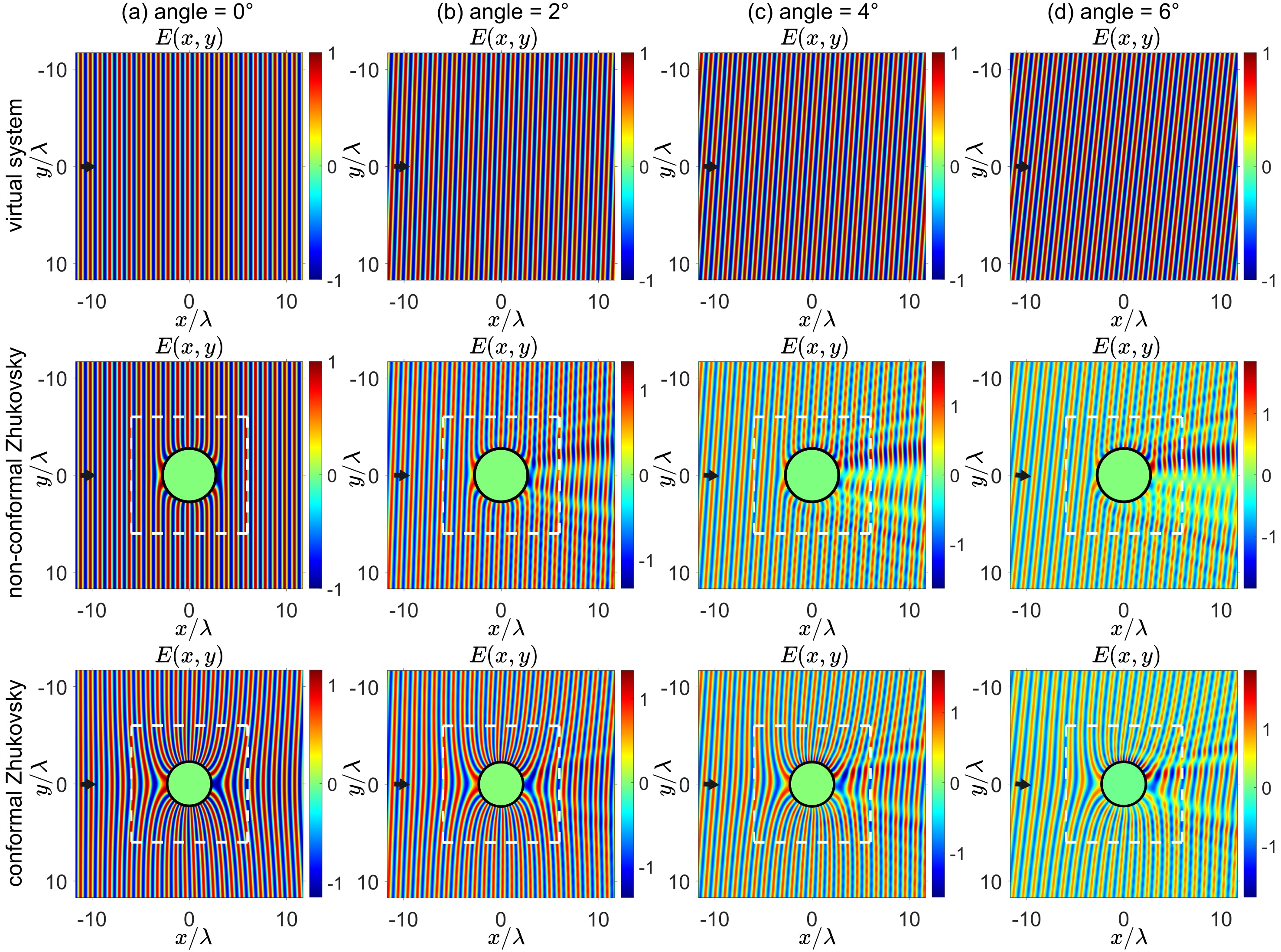}
\caption{Sensitivity of the cloaking efficiency on the input beam direction, for the homogeneous medium in the virtual system (first row), and the non-conformal (second row) and conformal (third row) cloaks shown in Fig. 2 of the main text. The input angles are: (a) $0^\circ$, (b) $2^\circ$, (c) $4^\circ$ and (d) $6^\circ$, with respect to the normal. 
}\label{Fig:angscan1}
\end{figure} 
\begin{figure}[t!]
\centering
\includegraphics[clip,width=\columnwidth]{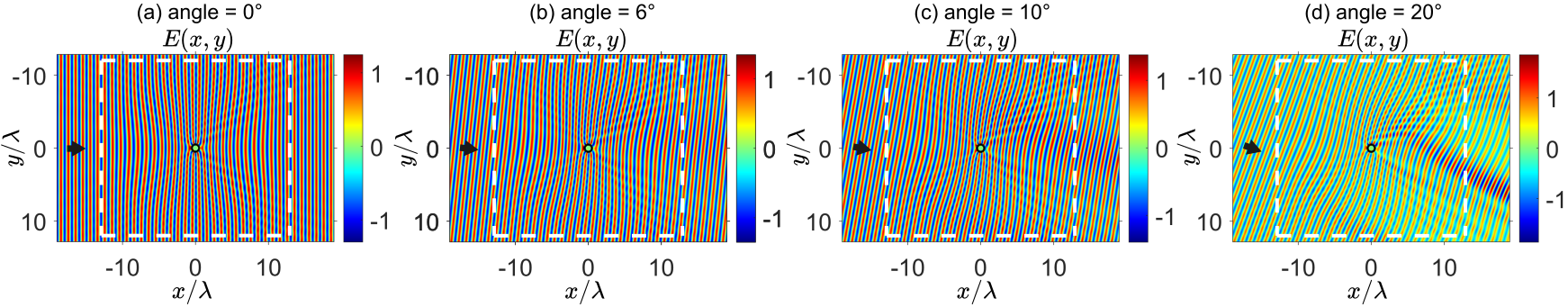}
\caption{Sensitivity of the cloaking efficiency on the input beam direction, for the cloak with the refractive index given in Fig. \ref{Fig:regime2}(a),(c). The input angles are: (a) $0^\circ$, (b) $6^\circ$, (c) $10^\circ$ and (d) $20^\circ$, with respect to the normal. 
}\label{Fig:angscan2}
\end{figure} 

\section{Sensitivity on the angle of incidence}
The cloak design with a refractive index given by the profile of Eq.~(4), assumes a transformation of the optical space for only a particular input wave—a plane wave with wavenumber $k_0$, that is normally incident on the medium. Moreover, our non-conformal Zhukovsky cloak is based on a tailored distribution of gain and loss, which is known to affect beams differently, depending on their direction of incidence. However, since the cloak efficiency exhibits robustness to changes of the input frequency, some robustness is also expected regarding changes on the direction of incidence. Note also that the conformal Zhukovsky cloak's efficiency is known to be sensitive to the incidence angle, so it is expected that this property will extend to its non-conformal modification. 

To test the input angle dependence of cloaking efficiency, we have scanned over various angles of incidence, meaning different incidence directions for incoming plane waves at the design frequency. The results for the refractive index of the conformal and non-conformal cases of Fig.~2 of the main text are shown in Figs.~\ref{Fig:angscan1}. It is visible from the results that both the conformal and non-conformal cases are, indeed, sensitive to the input wave direction, with the latter being slightly more so. A decreased sensitivity to the input beam direction is found for the refractive index of Fig.~\ref{Fig:regime2}(a),(c), as shown in Fig.~\ref{Fig:angscan2}(a),(c). 
\newpage

\bibliography{referencesnew}
\clearpage
\newpage

\end{document}